\documentclass[intlimits,twoside,a4paper]{article}

\usepackage[eqsecnum]{cmpj3}

\usepackage{bm}

\issue{2024}{27}{1}{13601}
\doinumber{10.5488/CMP.27.13601}

\title[Gibbs monolayers with different particle sizes]%
{Effects of charge and size on the coadsorption of counterionic colloids in Gibbs monolayers}
\author[J. M. G\'omez-Verd\'u, B. Mart\'inez-Haya, A. Cuetos]{J. M. Gómez-Verd\'u\orcid{0009-0004-3938-8126}, B. Mart\'inez-Haya\orcid{0000-0003-2682-3286}, 
	A. Cuetos\orcid{0000-0003-2170-0535}\thanks{Corresponding author: \email{acuemen@upo.es}}
}
\address{
Center for Nanoscience and Sustainable Technologies (CNATS), and Department
of Physical,
Chemical and Natural Systems, Pablo de Olavide University, Sevilla, Spain
}


\date{Received June 14, 2023, in final form July 30, 2023}

\begin{document}

\maketitle

\begin{abstract}
This study uses a coarse-grained Monte Carlo algorithm to model and simulate the coadsorption of a binary mixture of counterionic colloids in Gibbs monolayers. These monolayers form at a idealized air-water interface, with one non-soluble species confined at the interface and the second one partially soluble in the aqueous phase. The investigation focuses on the effect of colloidal size and charge on the thermodynamics and microstructure of the monolayer. We find that the composition of the monolayer evolves non-trivially with surface coverage, depending on the balance of steric and electrostatic forces. When the electrostatic interactions are weak, the soluble species is expelled from the monolayer upon compression, yielding a phase behaviour particularly sensitive to the relative size of the soluble and non-soluble colloids. By contrast, strong electrostatic interactions favour the stabilization of the soluble particles in the monolayer and the formation of quasi-equimolar fluids, with only a weak dependence on particle size. The combination of these phenomena results in the formation of a number of two-dimensional mesoscopic arrangements in the monolayer, ranging from diluted gas-phase behaviour to domains of aggregates and percolates, and to incipient crystalline structures. 
%
%
\printkeywords
\end{abstract}

\section{Introduction}

Computer simulation has consolidated as a fundamental tool for the rationalization and technological exploitation of the properties of molecular liquids, soft condensed matter, and materials in general. An important step in using computer simulation tools to investigate the properties of any system is to obtain a suitable model for the problem at hand. There is no single recipe for this task, and the level of complexity of the model to be used will depend on the properties and time and length scales to be studied. Very detailed microscopic models may provide quantitative information about some properties, but they usually make it difficult to explore the general physical characteristics that guide the behaviour of the system. This latter information is typically more accessible with simplified models, which may however compromise the acquisition of reliable quantitative and atomistic information. The choice and design of appropriate models for a specific study is an art in itself, which Professor Jaroslav Ilnytskyi has excelled throughout his career, in particular within the framework of complex molecular fluids \cite{ILN01,HUG08}. 

We present a computer simulation study inspired by the work of  prof. Ilnytsky, of the thermo\-dyna\-mics and structural properties of a multicomponent fluid of charged particles dispersed in an interfacial monolayer. More specifically, we investigate generic cases of Gibbs monolayers representing dispersed amphiphilic particles at an air-water interface \cite{MYE99,BIN17}. These systems began to be studied in modern times already in the 18th century, from the celebrated observations of films of olive oil on water carried out by Benjamin Franklin \cite{TAN89} and later revisited by Lord Rayleigh \cite{RAY91}. Subsequently, Agnes Pockels~\cite{RAY91,Poskels1892} designed the famous trough to measure the surface tension of monolayers, which was eventually improved by Irving Langmuir and his assistant Katherine Blodgett, implementing the efficient transfer of the monolayer to a solid substrate \cite{LAN20}. A monolayer is said to be of Langmuir-type if all the molecules are completely insoluble, while it is said to be a Gibbs monolayer if some of its constituents are partially soluble in the bulk of at least one of the phases that share the interface. In a general situation, the species belonging to the insoluble fraction will remain confined at the interface, while those belonging to the soluble fraction will be distributed between the fluid phases and the interface \cite{MEL98}. Gibbs monolayers find a broad range of applications, often related to sensing devices, where interfacial changes are induced when species in solution selectively bind to non-soluble colloidal receptors hosted by the monolayer \cite{ARI13,CHE12}. A nice example of this type of application is the interfacial binding of biomacromolecules, such as DNA fragments, by cationic surfactants. For such systems, it has been shown that the compression of the monolayer affects the concentration of the soluble biomolecules in the monolayer, a thermodynamic effect that is not yet well understood \cite{CHE12,XUE07}. Gibbs monolayers have also a potential to be utilized in the synthesis of thin-film topologies, facilitated by counterionic template frameworks that are incorporated to the interface from solution. An example of this is the fabrication of Langmuir-Blodgett layered structures based on guanidinium surfactants employing soluble carboxylate and phosphate moieties \cite{ARI13}.

Not surprisingly, Gibbs monolayers have attracted much attention and have motivated important developments from the field of molecular and colloidal simulation. In this context, different levels of approximation have been proposed, ranging from models with a highly detailed atomistic description of the molecules and particles involved, including water \cite{CAS10,WAN16,BAI22}, to coarse-grained models where entire parts of the molecules are modelled as beads, using various effective force fields \cite{DUC01,GON15}. In many studies, the monolayer is examined as a three-dimensional system, simultaneously modelling and simulating the interface and the two coexisting phases. In this way, it is possible to obtain valuable insights into the microscopic properties of the system, especially into the conformations that favour colloidal interactions~\cite{CAS10,GON15,BAI22}. However, due to the high computational cost, these models are usually restricted to small systems and short timescales, which often precludes the access to thermodynamic and mesoscopic structural properties.

In an attempt to complement the information obtained with more detailed modelling studies, we have recently proposed a simple coarse-grained model for the study of Gibbs monolayers \cite{CUE20}. Within our approach, only the interface itself is simulated as a two-dimensional system with interacting particles represented as flat disks. Amphiphilic and soluble particles can diffuse within the interface, but only the latter ones are capable of moving in and out of the interface. The interchange of soluble particles between the interface and the bulk phase is simulated with Grand Canonical Monte Carlo moves. Another aspect of our model  is that the interaction between charged particles is modelled as a dipolar interaction, an assumption that  resembles the different screening of colloidal charge induced by the two immiscible phases that conform the interface, e.g.,  the air-water interfaces \cite{HURD85,BLE13}. With these ingredients, we recently studied the phase behaviour and the structural properties of monolayers of mutually attractive soluble and non-soluble colloids  with the same size and absolute dipolar charge (but opposite sign)~\cite{CUE20}. For that system, an interesting sequence of disorder fluid, cluster fluid, and crystalline phases was found, confirming the potential richness of the structural landscape of Gibbs monolayers. In this article we extend the investigation by introducing asymmetry in colloidal size in the two complementary cases where either the soluble or the non-soluble component of the mixture is larger in size. In this way, we intend to isolate the size effects on interfacial behaviour from those of other colloidal properties.

\section{Model and methods}

The colloidal model and the Monte Carlo simulation methodology are similar to those employed in our previous work \cite{CUE20}. The Gibbs monolayer at the air-water interface is modelled using a simplified coarse-grained approach, which represents the system as a two-dimensional fluid, comprising a binary mixture of disk particles with different diameters, interacting through both steric and electrostatic interactions. One of the colloidal species (hereafter denoted as $S$) is assumed to be soluble, while its counterionic counterpart (denoted as $M$) is non-soluble (or amphiphilic) and it is therefore confined at the interface. Though the bulk of the liquid phase is not explicitly considered, it is taken into account on the assumption that it is in thermodynamic equilibrium with the interface. Thus, it is assumed that the chemical potential of the soluble $S$ particles, denoted as $\mu_S$, is the same in the bulk and at the interface. This chemical potential is dependent on the concentration of the particles
in solution, $C_S$, through the expression $\mu_S = \mu^{\text{ref}}_S + k_{\text B} T \ln \gamma C_S$, where $\mu^{\text{ref}}_S$ is the chemical potential in a reference state and $\gamma$ is the activity coefficient ($\gamma = 1$ for an ideal solution). Here, $k_{\text B}$ is the Boltzmann constant, and $T$ is the temperature. The chemical potential modulates the incorporation of $S$ particles from the solution to the monolayer and this is emulated by means of Monte Carlo simulations in the $\mu_sN_M\Pi T$ ensemble, where $N_M$ denotes the fixed number of non-soluble particles in the monolayer and $\Pi$ is the surface pressure. The Monte Carlo algorithm samples the equilibrium configurations with a combination of random displacements of $S$ or $M$ particles, with the insertion or removal of $S$ particles, and changes in the surface area of the monolayers. The acceptance or rejection of these move attempts will depend on the interaction between the particles, the chemical potential $\mu_S$ and the surface pressure for volume changes, following general acceptance rules based on the Metropolis algorithm \cite{FRE02}. 

The simulations were run on systems of $N_{M}=$ 4000 non-soluble $M$ colloids, while the number of soluble $S$ colloids is allowed to change throughout the simulation without constraints. Each Monte Carlo cycle involved $N_M$ attempts for random displacements of a particle ($S$ or $M$), for changes in the surface area, and for the insertion or deletion of a particle $S$ in or from the interface. Using the values optimized in our previous work, the displacement of a random particle was attempted with frequency $w_d=0.765$, the changes of surface with frequency $w_a=0.0005$, while the insertion/deletion moves are both attempted with frequency $w_i=0.230$. A typical simulation at a given surface pressure and chemical potential starts from a random initial configuration of low surface density. To equilibrate the system, about $1-2\cdot10^6$ Monte Carlo cycles were applied. Additional $5\cdot10^5$ Monte Carlo cycles were then run to obtain the averages of the observables of interest.

The choice of the interaction potential between the particles is a key aspect of the model. The interaction between charged particles at an air-water interface is a far from trivial problem that has attracted the interest of many researchers \cite{HURD85,MAJ18,CHE09,LIA16,BLE13}. A long-range $r^{-3}$ dipole-dipole interaction has been proposed to describe the interaction between charged interfacial particles, as a consequence of different shielding of the Coulombic forces in the air and water sides of the interface. Here, we consider such dipolar interaction, on top of a short-range steric repulsion between the particles. Other types of interaction, such as capillary interactions, have been pointed out as relevant \cite{BLE13}, but are not considered in our model. With all these ingredients, the interfacial colloids interact through a pair potential $U(r) = U_{LJ}(r)+U_D(r)$, expressed as the sum of short-range steric repulsions and long-range effective dipole interactions \cite{CUE20}.

On the one hand, the short-range interaction is represented by a truncated and shifted Lennard-Jones potentials of the form

\begin{equation}\label{eq1}
U_{LJ} = \left\{ \begin{array}{cc}
4 \epsilon \left[  \left( \sigma_{ij}/r \right)^{12} -
\left( \sigma_{ij}/r \right)^6 + 1/4 \right] & ~~   r \leqslant \sqrt[6]{2}\,\sigma_{ij},
\\
0 & ~~ r > \sqrt[6]{2}\,\sigma_{ij}. \end{array} \right.
\end{equation}
Here, $\sigma_{ij}=0.5(\sigma_i+\sigma_j)$ is the contact distance between the interacting particles, with diameters $\sigma_i$ and $\sigma_j$, respectively. $r$ is the distance between particles. Note that since the potential is truncated at the minimum of the Lennard-Jones potential well, it is a purely repulsive contribution to the net interaction. 

On the other hand, the dipolar contribution to the interaction potential reads as follows

\begin{equation}\label{eq2}
U_{D} = \left\{ \begin{array}{cc}
\epsilon \Delta_i\Delta_j\left[(\frac{\sigma}{r})^3-(\frac{\sigma}{r_s})^3\right]& ~~   r \leqslant r_s,
\\
0 & ~~ r> r_s,\ \end{array} \right.
\end{equation}
 where $\sigma$ and $\epsilon$ are the unit of length and energy, respectively. $r_s$ is the cutoff distance of the dipole interaction, which as in our previous article is set as $r_s=30\sigma$. $\Delta_i$ and $\Delta_j$ are the dimensionless dipolar charge of particles $i$ and $j$. The values assigned to these dipole moments in this study are $3$, $6$ or $9$ in reduced units, as commented below.

\section{Results}\label{results}

In our previous work \cite{CUE20}, we focused on the reference situation of $M$ and $S$ colloidal particles with equal dipolar charge modulus and diameter. Here, we consider situations in which the colloidal particles $M$ and $S$ are asymmetric in size.  Two complementary cases are addressed in which either the non-soluble $M$ particles or, alternatively, the soluble $S$ particles are larger in size. We  specifically considered the case with $\sigma_M=\sigma$ and $\sigma_S=3\sigma$, as well as the reverse situation where $\sigma_M=3\sigma$ and $\sigma_S=\sigma$.  The temperature and the chemical potential of the soluble particles were set to $T^*=k_{\text B}T/\epsilon=1$ and $\mu^* = \mu/\epsilon=-8$, respectively, while surface pressure was varied in the range $\Pi^* = \Pi\sigma^2/\epsilon =0.001-3$. Our previous study showed that within this range of parameters the Gibbs monolayers display a rich phase diagram.

\subsection{Thermodynamic study of the monolayers}

\begin{figure}[!t]
	\centerline{\includegraphics[width=1.0\textwidth]{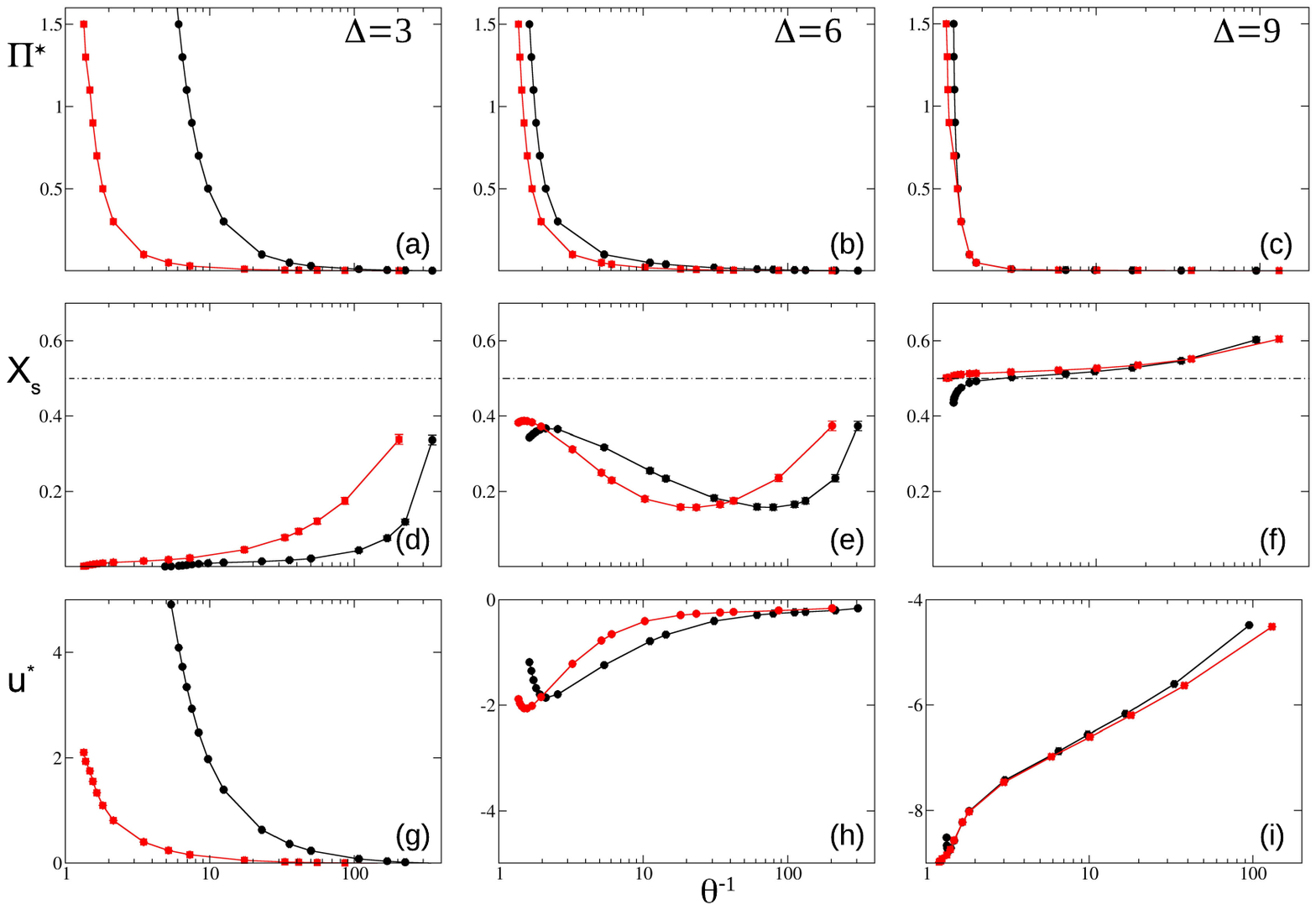}}
	\caption{(Colour online) Monte Carlo isotherms for systems with colloids of diameters $\sigma_S=3,\sigma_M=1$ (black lines and circles) and $\sigma_S=1,\sigma_M=3$ (red lines and circles) and dipole moments $\Delta=3$ (panels a, d and g), $6$ (panels b, e and h) and $9$ (panels c, f and i). The top (a to c), middle (d to f) and bottom (g to i) panels plot the surface pressure $\Pi^* = \Pi\sigma^2/\epsilon$, the mole fraction of $S$ colloids $X_s$ and the energy per particle $u^*=U/N\epsilon$, respectively, as a function of the inverse of the surface coverage, $\theta^{-1}$. The horizontal dash-dotted lines indicate the equimolarity of the oppositely charged colloids ($X_s=0.5$).}
	\label{fig1}
\end{figure}

The equations of state of the systems under study are characterized in figure~\ref{fig1}, in terms of the evolution of the surface pressure $\Pi^* = \Pi\sigma^2/\epsilon$, the molar fraction of soluble colloids $X_s$ and the internal energy per particle in reduced units, $u^*=U/N\epsilon$,  as a function of the inverse coverage, $\theta^{-1} = A/A_p$. Note that $\theta^{-1}$ represents the interfacial area per particle; $A$ is the total area of the interface, while $A_p=N_M\cdot a_M + N_S\cdot a_S$ is the effective area physically occupied by the particles at the interface, with $a_i=\piup\sigma^2_i/4$ the area of a particle of species $i$. 

The main finding of the present study is that for weak dipolar interactions ($\Delta = 3$), the monolayer tends to be equimolar in the two colloidal components in the high dilution limit ($X_s\approx X_m\approx 0.5$, as $\theta^{-1}\rightarrow\infty$), whereas it becomes progressively enriched in the non-soluble component $M$ upon compression. The soluble component $S$ tends to be expelled from the monolayer, so that at high coverage the monolayer becomes essentially a monocomponent system of $M$ colloids ($X_m\rightarrow$1, $X_s\rightarrow 0$, as $\theta^{-1}\rightarrow0$, see figure~\ref{fig1}d).  A consequence of this behaviour is that the system displays a marked asymmetry with respect to the size ratio of the $M$ and $S$ colloids. Figures~\ref{fig1}a and \ref{fig1}g show that when the smaller colloid is the insoluble component ($\sigma_M=1$), the monolayer displays a steep increase in pressure and internal energy with increasing coverage, at $\theta^{-1}<10$. A similar increase occurs at much higher coverages, onset at $\theta^{-1}<2$, when the insoluble colloid is the larger one ($\sigma_M=3$). In general terms, at any given surface coverage away from the dilution limit, much higher pressures and internal energies are associated with the $\sigma_M=1$ fluid in comparison to the $\sigma_M=3$  fluid.

The exclusion of the $S$ particles  at high coverage just described for $\Delta = 3$, can be rationalized in terms of an increase in entropy associated with the larger area available to the intefacial particles. At a higher dipolar charge, such entropic effects are balanced by the attractive interaction between $S$ and $M$ particles, which favours the presence of $S$ colloids in the monolayer. Hence, the thermodynamic behavior of the monolayer changes qualitatively as the dipolar interactions become significant. This also brings the system to a much more symmetric dependence on colloid size ratio, meaning that it becomes progressively less relevant whether the larger colloids belong to the soluble  component or to the insoluble one. For the present study, the isotherms with $\sigma_M=3$ ($\sigma_S=1$) and  $\sigma_M=1$ ($\sigma_S=3$) are similar for  $\Delta = 6$ and become almost overlapping for  $\Delta = 9$ (see figure~\ref{fig1}). Interestingly, the strong dipole interactions largely favour the compression of the monolayer, which proceeds with small changes in pressure and with an internal energy $u^*$, that remains negative throughout the full isotherms and becomes more negative upon compression. Note that this is in strong contrast with the increasingly net positive $u^*$ values attained for $\Delta = 3$. Negative values of $u^*$ are associated with dominant attractive interactions, indicating that an appreciable fraction of $S$ colloids remains at the interface at all surface coverages. 

For $\Delta = 6$ at intermediate dilutions ($\theta^{-1}>10$), $u^*$ is weakly dependent on the surface coverage  [figure~\ref{fig1}(h)]. In this range, entropic effects dominate, and the $S$ particles are still sterically ejected from the system upon compression, yielding a decrease in $X_s$ qualitatively similar to that found for $\Delta = 3$ [figure~\ref{fig1}(e)]. Interestingly however, the $S$-particles re-enter the monolayer at sufficently high surface coverages, correlating with a steep decrease in the internal energy. In the limit of high coverage, the molar fraction reaches the values similar to those obtained in the high dilution regime, $X_s\approx0.4$. For the strongly interacting $\Delta=9$ monolayer, the molar fraction of $S$ colloids is larger than $0.5$  at high dilution, indicating that the incorporation of an excess of soluble colloids to the monolayer is favored under these conditions. The composition of the monolayer approaches then equimolarity $X_s\approx0.5$ at high packing fractions.

\subsection{Microscopic structure of the monolayers}

	\begin{figure}[!t]
	\centerline{\includegraphics[width=0.99\textwidth]{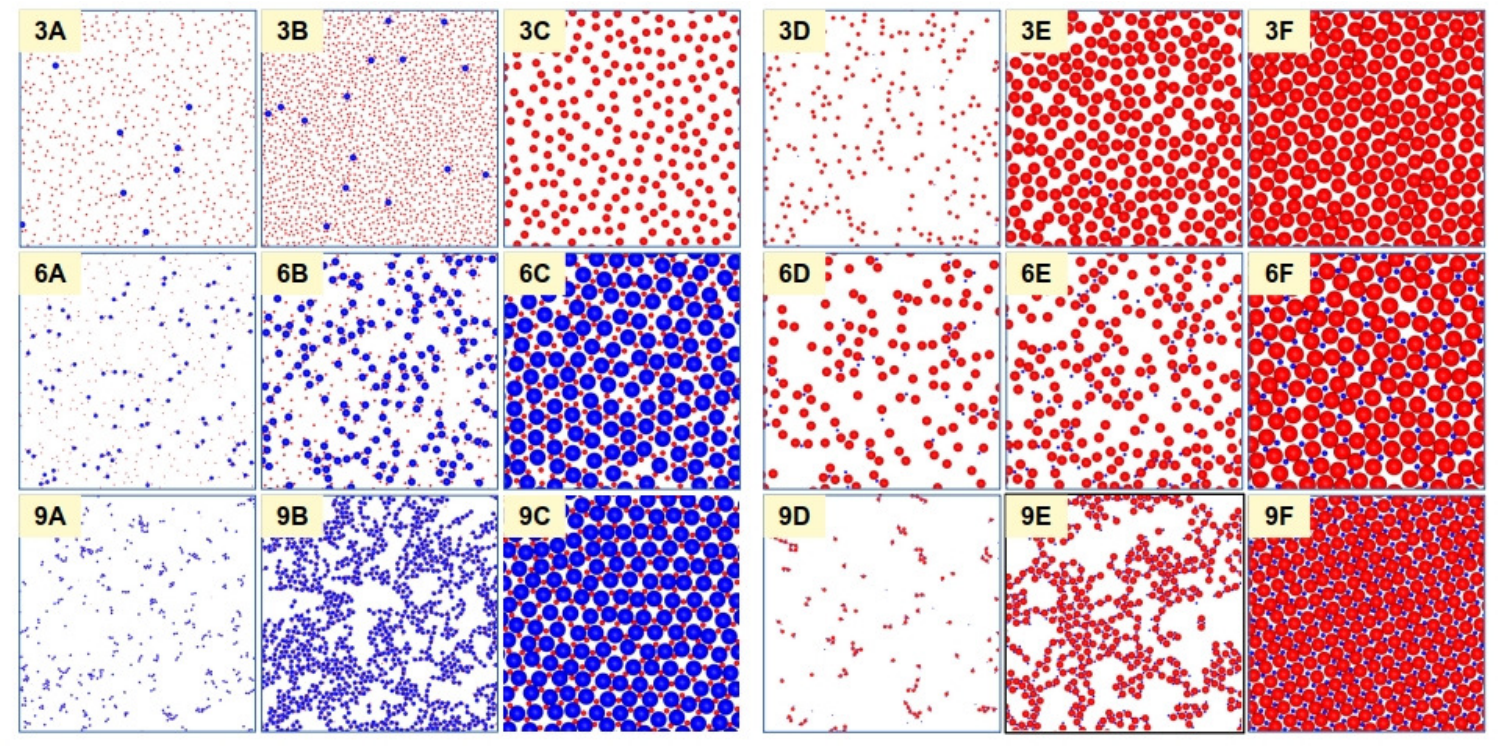}}
	\centerline{\includegraphics[width=0.65\textwidth]{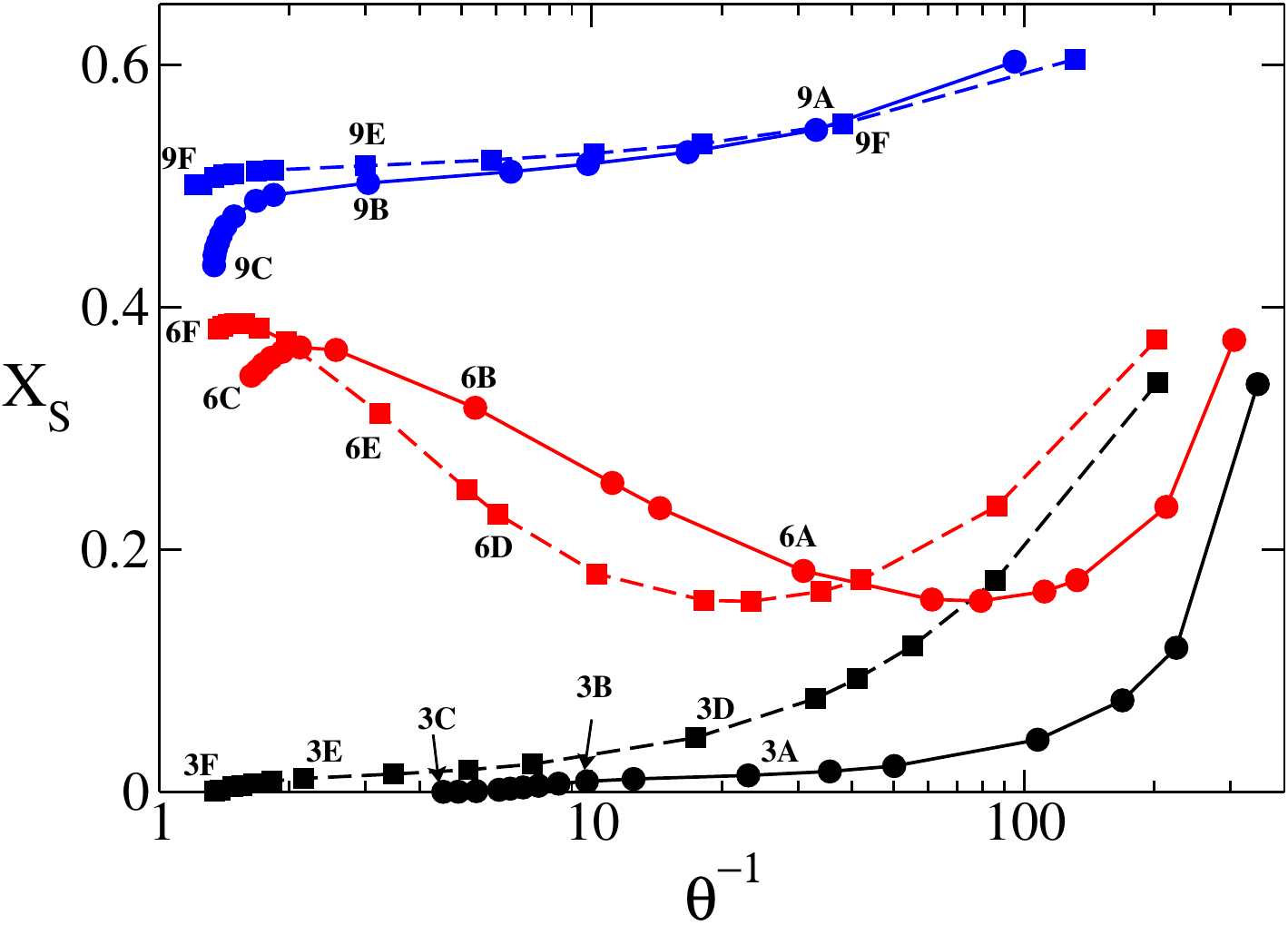}}
	\caption{(Colour online) Representative snapshots of microscopic configurations of
		the Gibbs monolayers explored in this study. Red and blue disks
		represent the nonsoluble ($M$) and the soluble ($S$) colloids,
		respectively. Snapshots with labels A, B and C correspond to situations where $\sigma_M=1$ and $\sigma_S=3$, while labels D, E and F indicate the reciprocal cases, with $\sigma_M=3$ and $\sigma_S=1$. The number in the labels indicates the dimensionless dipolar charge in each case, $\Delta=3$, $6$ or $9$. The composition of the monolayer in each state can be inferred from the representation of $X_s$ in the bottom panel, which shows an extract of the data included in panels (c), (e) and (f) of figure~\ref{fig1}.}
	\label{fig2}
\end{figure}

The thermodynamic properties described so far are intimately related to the microscopic and mesoscopic structures adopted by the particles within the monolayer. The topology and relative stability of  different interfacial structural arrangements are strongly conditioned by the incorporation of soluble colloids $S$ from the bulk phase. Figure \ref{fig2} provides an overview of different structures observed in our work. The ejection of S colloids from the monolayer at $\Delta=3$ can be appreciated in the sequences of configurations from 3A to 3C ($\sigma_M=1$, $\sigma_S=3$) and from 3D to 3F ($\sigma_M=3$, $\sigma_S=1$). In the high density configurations 3C and 3F, the monolayer consists of a monocomponent fluid of $M$ colloids. A trend towards the hexatic crystalline order characteristic of the two-dimensional discotic particle fluid is observed at high pressures and surface coverages (configuration 3F), although a consolidated crystalline phase is not found within the thermodynamic range included in our study for $\Delta=3$. When the dipolar charge is increased, the particle aggregation and cluster formation tend to occur even in the high dilution limit. This phenomenon is particularly pronounced for $\Delta=9$, (configurations 9A and 9D) where linear aggregates of several particles are observed, whereas for $\Delta=6$ (panels 6A and 6D) mainly $M-S$ dimers are found. As the concentration increases, the formation of clusters is more evident and percolated arrangements are formed. In the more dense monolayers of the strongly interacting colloids, the percolates colapse to an incipient square crystalline arrangements. 
\begin{figure}[!t]
	\centerline{\includegraphics[width=1\textwidth]{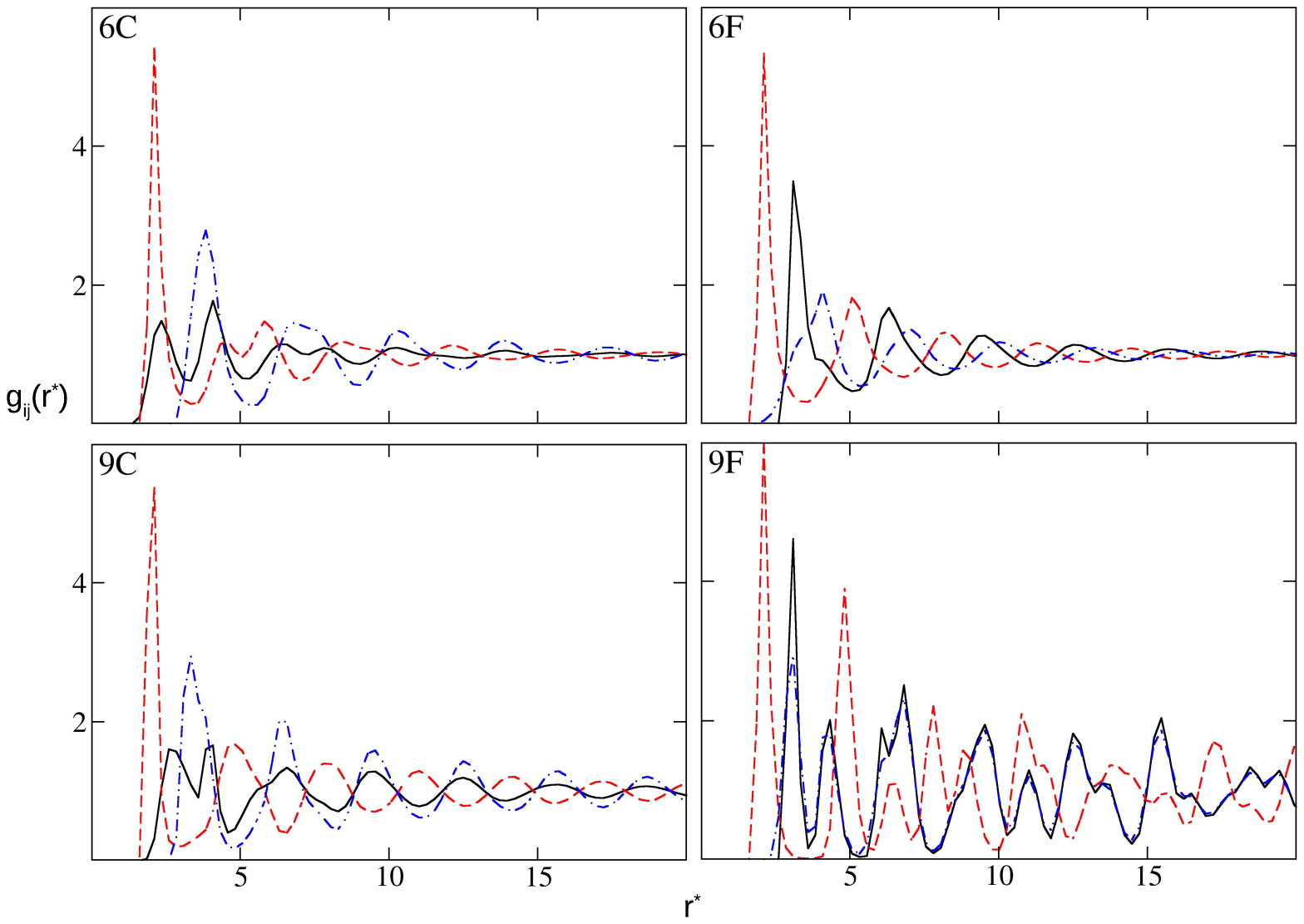}}
	\caption{(Colour online) Radial functions associated with the $M-M$ ($g_{MM}$) and $S-S$ ($g_{SS}$) self-correlations between  alike-charged colloids (black solid and blue dotted dashed curves, respectively) and with the $M-S$ cross-correlations ($g_{MS}$) between oppositely charged colloids (red dashed curves) for the states 6C (top left), 6F (top right), 9C (bottom left) and 9F (bottom right) described in figure \ref{fig2}. Note that in 6C and 9C cases $\sigma_M=1$ and $\sigma_S=3$, while for 6F ad 9F $\sigma_M=3$ and $\sigma_S=1$.}
	\label{fig3}
\end{figure}
Despite the abundance of structural defects, we obtain the evidence for square cristalline symmetry in the high coverage limit (e.g. configuration 9F in figure~\ref{fig2}). The analysis of the corresponding radial distribution functions confirms these findings. Figure~\ref{fig3} depicts an illustrative set of radial $M-M$, $S-S$ and $M-S$ correlation functions ($g_{MM}$, $g_{SS}$ and $g_{MS}$ respectively). Configuration 9F is representative of the formation of a crystalline arrangement; the position of the peaks in the correlation functions indicates that the structure corresponds to a square lattice, in which the smaller colloids are located in the intersticial sites generated by the larger colloids. The distribution functions for configurations  6C, 6F and 9C display a strong correlation at short and medium distances, resembling the domains of crystalline structure, which however does not survive at long range. These configurations deserve further investigation to ensure that they correspond to true equilibrium states and are not trapped in a metastable configuration. Note that our simulations were performed by compressing the monolayer from the high dilution limit. Metastability could in particular explain the sharp drop in $X_S$ observed for the $\Delta= 6$ and $9$ cases at high coverage, and the differences in the equation of state observed in this range while comparing the monolayers qith $\sigma_M=3$ {vs.} $\sigma_M=1$ (figure~\ref{fig1}).

\begin{figure}[!t]
	\centerline{\includegraphics[width=0.60\textwidth]{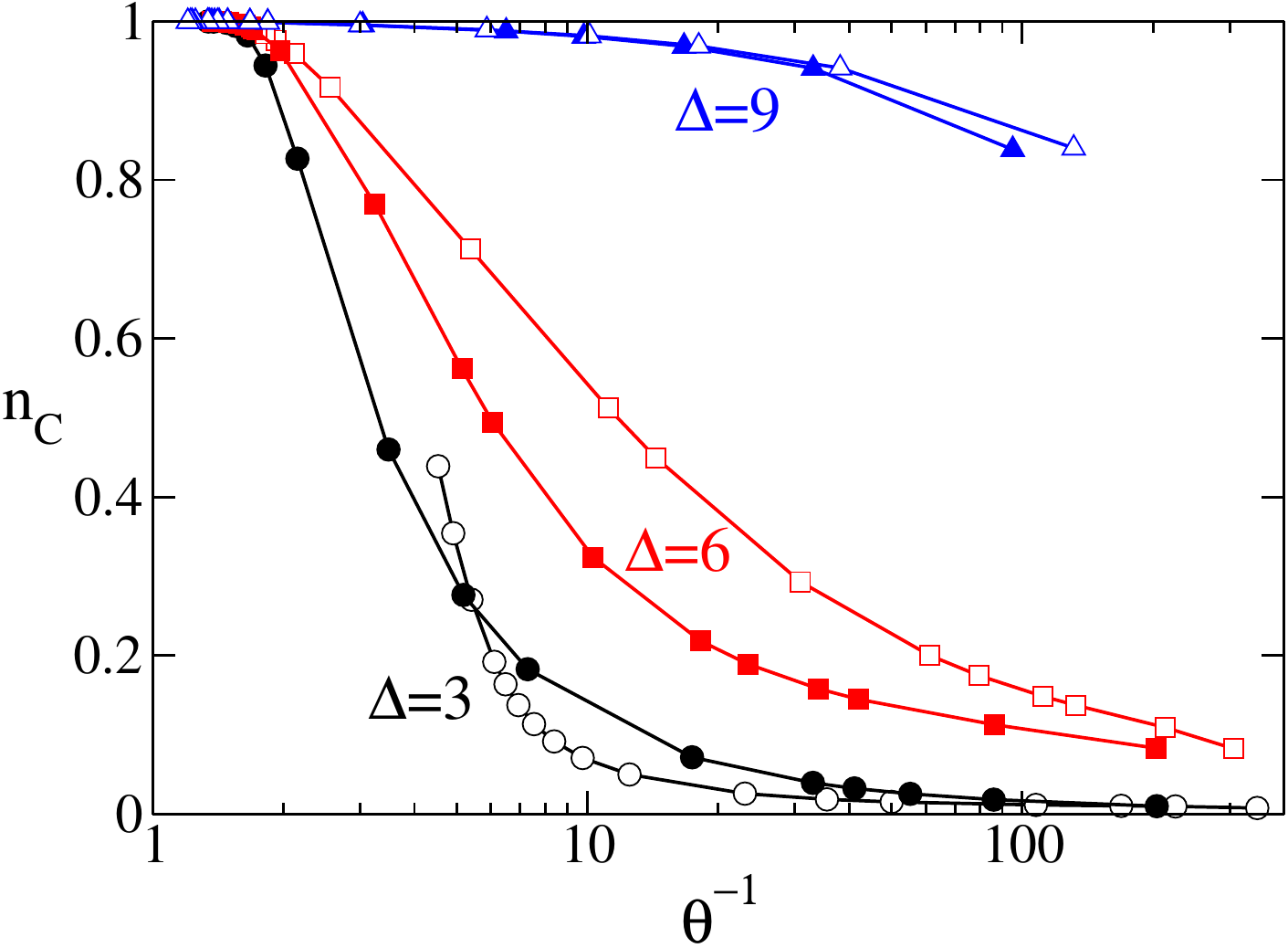}}
	\caption{(Colour online) Fraction of colloids in clusters $n_C$ as a function of surface coverages $\theta^{-1}$, for $\Delta=3$ (black lines and circles), $6$ (red lines and squares) and $9$ (blue lines and triangles). Open symbols indicate the cases with $\sigma_M=1$ and $\sigma_S=3$ while solid symbols are the reciprocal case where $\sigma_M=3$ and $\sigma_S=1$.}
	\label{fig4}
\end{figure}

Colloidal aggregation in the strongly interacting monolayers is intriguing. Clustering is actually a common phenomenon in the fluids of charged colloids at air-water interfaces, as known from experiments~\cite{SUZ13,NAL17}. The transition from a regime dominated by small aggregates to either mesoscopic aggregation or long-range percolated arrangements is subtle and challenging to predict. Here, we assess the extent of aggregation by analyzing the interconnection between neighbouring colloids as a function of surface coverage. To this end, we considered that two particles belong to the same cluster if the distance between them is smaller than $\sigma_{ij}+0.5\sigma$, where $\sigma_{ij}$ is the contact distance between the particles. Within this framework, two types of ensemble-averaged magnitudes were employed to characterize the clustering of particles. Firstly, the distribution, $F(i)=i\cdot C(i)/N_T$, represents the fraction of particles belonging to a cluster of size $i$, where $C(i)$ is the average number of clusters of size $i$ and $N_T$ is the average number of particles at the interface. Secondly, the clustering paremeter, $n_C$, represents the fraction of colloids embedded in clusters of any size, including dimers and larger clusters ($n_C =\sum_{i>1} F(i)$).  

\begin{figure}[!b]
	\centerline{\includegraphics[width=0.65\textwidth]{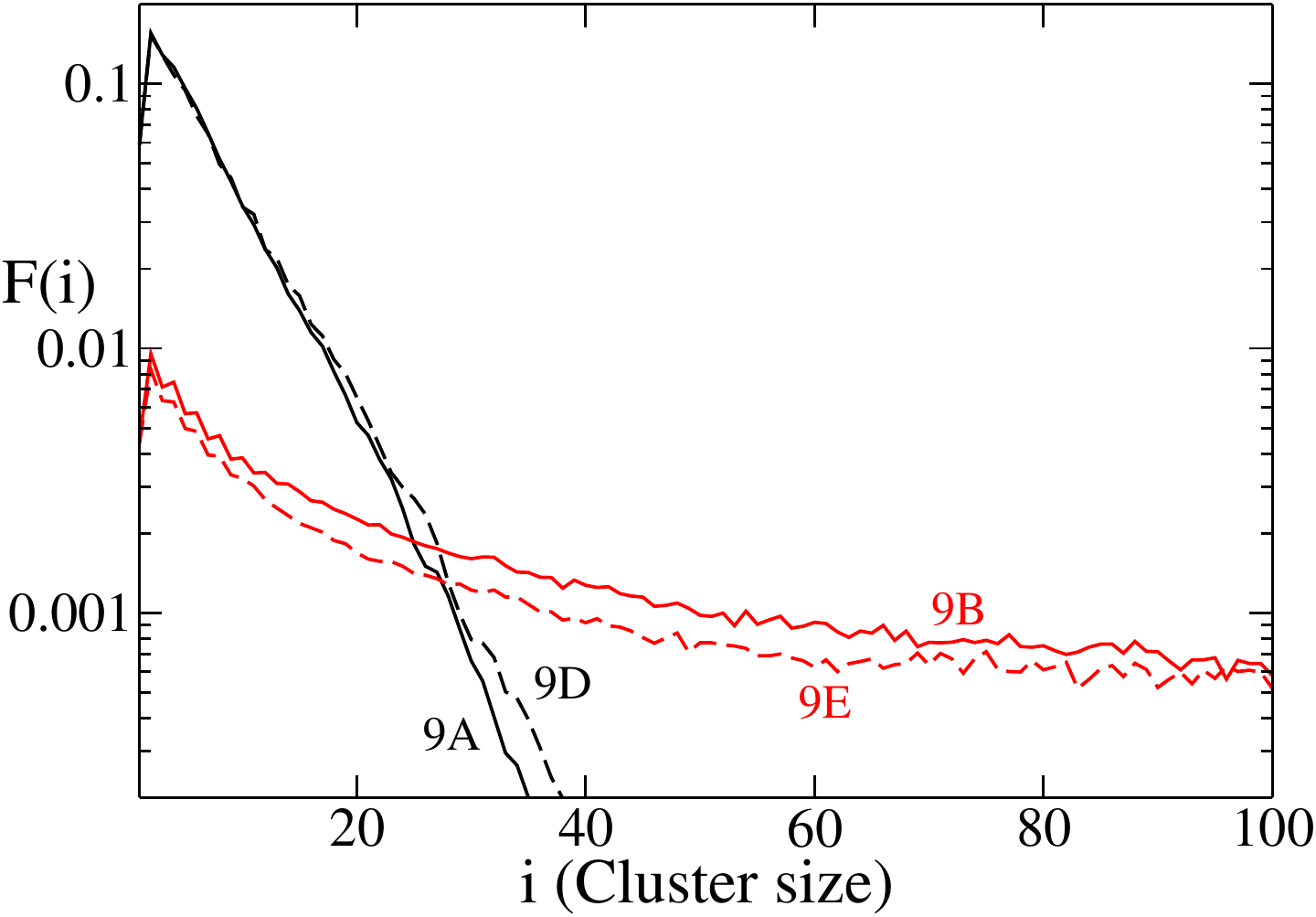}}
	\caption{(Colour online) Fraction of colloids in clusters of a given size $i$ for some states of the monolayer with $\Delta=9$ (see labels in figure~\ref{fig4}). Solid lines correspond to states with $\sigma_M=1$, while dashed lines correspond to states with $\sigma_M=3$.}
	\label{fig5}
\end{figure}

Figure \ref{fig4} depicts the dependence of $n_C$ on $\theta^{-1}$ for the fluids under study. Consistently, $n_C\rightarrow1$ at high coverages, and it decreases steadily as the system approaches the high dilution limit. Cluster dissociation as the area accessible to the colloids increases, mainly driven by entropy, is clearly apparent for $\Delta =3$, with $n_C$ reaching the values close to 0 at high dilution. For stronger dipolar interactions, clustering is still noticeable in the expanded monolayers. In the case of $\Delta=6$, even for $\theta^{-1}>$\,100, around 10\% of the colloidal particles are incorporated to a cluster. For $\Delta=9$, the particle aggregation at high dilution still reaches remarkable values for $n_C$ above 80\%. The domains of clustering and percolation are characterized in greater detail in figure~\ref{fig5}, which shows the cluster distributions $F(i)$ for relevant configurations of $\Delta=9$ monolayer. We find that colloids are distributed in clusters with a broad distribution of sizes. It can be noted that even at high dilution (configurations 9A and 9D), the fraction of isolated colloid monomers remains below 10\%, while dimers amount to almost 20\%, with the majority of the remaining 70\% corresponding to clusters with sizes up to $i\approx20$. In the percolation domain (configurations 9B and 9E) not more than 0.5\% of the colloids stay as monomers  and the cluster distribution broadens significantly, extending to sizes $i\approx 50-100$. Notably, a coincident aggregation behaviour is found for the monolayers with ($\sigma_M=1$, $\sigma_S=3$) and with ($\sigma_M=3$, $\sigma_S=1$) when cluster distributions are compared at similar surface coverages (9A {vs.} 9D and 9A  vs. 9D, in figure~\ref{fig5}). These results confirm the idea that for intermediate and high values of $\Delta$, the presence of aggregates is one of the main characteristics of Gibbs monolayers. Similar results were obtained for colloidal particles of equal size~\cite{CUE20}. One question we leave open for future work is the stability of the clusters, whether they have  long lifetimes or, on the contrary, they are continuously forming and fracturing. Answering this question demands simulation techniques other than the Monte Carlo algorithm used here, such as molecular dynamics simulations.

 \newpage
\section{Conclusions and final remarks}

This work presents a computer simulation study of the structural and thermodynamic behaviour of Gibbs monolayers in the case where the soluble and non-soluble colloids have different diameters. As such, it is an extension of a previous work devoted to the case of colloids with equal size \cite{CUE20}. This study also corroborates the diversity of phases and structures that are possible in Gibbs monolayers depending on colloidal parameters such as size and dipole charge, as well as thermodynamic conditions. Most of these phases and structures consist of disordered fluids with the particles forming aggregates. 

It is interesting to see how the two complementary situations that we have studied, namely cases in which the non-soluble colloids $M$ are smaller ($\sigma_M=1$ and $\sigma_S=3$), or larger ($\sigma_M=3$ and $\sigma_S=1$), display the most marked differences with each other in the case of weak colloid-colloid interactions (represented here by a dipolar moment $\Delta=3$). This finding can be traced back to the fact that the soluble colloids $S$ are expelled from the monolayer with increasingly surface coverage, eventually leading to an interfacial a monocomponent fluid of $M$ particles. Stronger colloid interactions ($\Delta$=\,$6$, $9$) promote the stabilization of $S$ colloids at the interface, leading to comparable fractions of soluble and non-soluble colloids at the monolayer. Under these conditions, most of
these phases and structures consist of disordered fluids with the particles forming aggregates. In very dilute systems we find gas
phases of free particles and dimers.  In denser systems, progressively larger clusters are formed, eventually leading to percolated arrangements. In the limit of high coverages, we find evidence of square crystalline structures, but also of fluids with short-range order which, due to the presence of defects, the correlation between particles is lost at long distances.

In short, in this work we have  confirmed the potential complexity of Gibbs monolayers as a function of the system parameters, and the capacity of the outlined simulation methodology to explore it. We believe that the results obtained may be useful for a better understanding of the design and tailoring of this type of two-dimensional fluids for specific technological applications.

\section*{Acknowledgments}

The authors acknowledge support from Consejer\'ia de Transformaci\'on Económica, Industria, Conocimiento y Universidades de la Junta de Andaluc\'ia/FEDER (projects P20-00816 and P20-01258), and from the Spanish Ministerio de Ciencia e Innovación and FEDER (Projects no. PID2021-126121NB-I00,  PID2019-110430GB-C22 and TED2021-130683B-C21). We are thankful to C3UPO for the HPC facilities provided.


\bibliographystyle{cmpj}
\bibliography{biblio-layers}

\ukrainianpart

\title{Вплив заряду та розміру на коадсорбцію протиіонних колоїдів у моношарах Гіббса}
\author[Ж. М. Гомес-Верду, Б. Мартінес-Айя, А. Куетос]{Ж. М. Гомес-Верду, Б. Мартінес-Айя, А. Куетос}

\address{
Факультет фізичних, хімічних та природніх систем, Університет Пабло де Олавіде, 41013 Севілья, Іспанія
}

\makeukrtitle

\begin{abstract}
	У цьому дослідженні застосовується огрублений алгоритм Монте-Карло для моделювання коадсорбції бінарної суміші протиіонних колоїдів у моношарах Гіббса. Моношари утворюються на ідеалізованій межі розділу ``повітря-вода'', при цьому нерозчинна речовина обмежена межею розділу, а друга речовина част\-ко\-во розчинна у водній фазі. Дослідження закцентовано на вивченні впливу розміру колоїдної частинки та її заряду на термодинаміку та мікроструктуру моношару. З'ясовано, що склад моношару міняється нетривіальним чином зі зміною покриття поверхні в залежності від балансу стеричних та електростатичних сил. Коли електростатичні взаємодії слабкі, розчинна речовина викидається з моношару під час його стиснення, формуючи фазову поведінку, яка особливо чутлива до відношення розмірів розчинних і нерозчинних колоїдів. Навпаки, сильні електростатичні взаємодії сприяють стабілізації розчинних частинок у моношарі та утворенню квазіеквімолярних плинів з незначною залежністю від розміру частинок. Поєднання цих явищ призводить до утворення низки двовимірних мезоскопічних структур у моношарі в діапазоні від розрідженої газової фази до областей, що складаються з агрегатів та перколятів, і далі аж до зародків кристалічних структур.
	\keywords моношари Гіббса, метод Монте-Карло, рівняння стану, агрегація
\end{abstract}

  \end{document}